\documentclass[prc,aps,nofootinbib,twocolumn,superscriptaddress]{revtex4-1}   
\usepackage{epsfig}  
\usepackage{graphicx}  
\usepackage{amssymb}
\usepackage{amsmath}
\usepackage{color}  

\usepackage{dsfont}

\begin{document}

\title{ Examining empirical evidence of the effect of superfluidity on the fusion barrier }

\author{Guillaume Scamps}
\email{scamps@nucl.ph.tsukuba.ac.jp}
\affiliation{Center for Computational Sciences, 
University of Tsukuba, Tsukuba 305-8571, Japan}

%


%
%

\begin{abstract}
\begin{description}
\item[Background] 
Recent Time-Dependent Hartree-Fock-Bogoliubov (TDHFB)  calculations predict that the superfluidity enhances the fluctuations of the fusion barrier. This effect is not fully understood and not yet experimentally revealed .   
\item[Purpose]  
The goal of this study is to empirically investigate  the effect of the superfluidity on the distribution width of the fusion barrier.
\item[Method] 
Two new methods are proposed in the present study.
First, the local regression method is introduced and used to determine the barrier distribution. The second method that requires only the calculation of an integral of the cross section is developed to determine accurately the fluctuations of the barrier. This integral method, showing the best performance,  is systematically applied  to 115 fusion reactions.
\item[Results]  The fluctuations of the barrier for open-shell systems are on average larger than those for magic or semi-magic nuclei. This is due to the deformation  and the superfluidity. To disentangle these two effects, a comparison is made between the experimental width and the width estimated from a model that takes into account the tunneling, the deformation, and the vibration effect. This study reveals that the superfluidity enhances the fusion barrier width.
\item[Conclusions]  This analysis shows that the predicted effect of the superfluidity on the width of the barrier is real and is of the order of 1 MeV.
\end{description}
\end{abstract}

\maketitle

\section{Introduction}
\label{Sec:intro} 

Recently, new applications have been developed to study microscopically the reactions between superfluid nuclei \cite{Has16,Mag16}. Using the Time-Dependent Hartree-Fock-Bogoliubov (TDHFB) theory with a Gogny interaction,  the reaction $^{20}$O+$^{20}$O is simulated in Ref. \cite{Has16}.  It is shown in this reaction where both fragments are superfluid that the fusion barrier depends on the initial relative gauge angle. An amplitude of $\Delta B$=0.4 MeV is found between the maximum and minimum heights of the barrier. This difference is due to the pairing interaction between the two fragments that is either attractive or repulsive depending on the relative phase. This effect of the superfluidity is not taken into account in actual fusion model \cite{Bac14}.

For the heavier system $^{90}$Zr+$^{90}$Zr, Magierski et al. with the FaNDF0 functional without spin-orbit interaction find a very large amplitude of $\Delta B$=30 MeV. With the same type of calculation, for the reaction $^{44}$Ca+$^{44}$Ca, a value of $\Delta B$=2.3 MeV is found \cite{Sek17}.  This effect is also seen on $^{120}$Sn+$^{120}$Sn \cite{Bul17} and on asymmetric reactions $^{86}$Zr+$^{126}$Sn \cite{Sek17b}.

Nevertheless, those calculations assume a semi-classical treatment of the collective variables. Indeed, the gauge angles should not be treated as a parameter of the reaction. A more elaborate method is to restore the initial symmetry in both fragments using a projection technique. A first attempt to restore the symmetry in TDHFB has been achieved recently with simplifying assumptions \cite{Sca17} to study the Josephson effect, but this method can not be directly used to determine the fusion barrier. 

A simpler method to restore the symmetry is proposed in \cite{Sca17_proc,Reg17}. It  assumes an initial uniform distribution of relative gauge angles. Then, from this distribution, an ensemble of independent TDHFB trajectory is performed leading to a final distribution of the observable of interest. In a toy model, comparisons to the exact solution show that  the first and second moments of the semi-classical TDHFB distributions are accurate with respect to the exact distributions. Hence, it is expected that the TDHFB may reproduce the standard deviation of the barrier distributions. However, it has to be kept in mind that the TDHFB method cannot reproduce the tunneling effect that would increase the fluctuations of the barrier distribution. More complex methods could solve the problem with a simultaneous description of the tunneling effect and the superfluidity.
For example, the Density-constraint-TDHFB method (that remains to be developed) based on the  Density-constraint-Time-Dependent Hartree-Fock theory \cite{Uma06} with the consideration of the pairing correlations. In the absence of more complex theory, one can still consider that the fluctuations of the barrier due to the pairing gauge angle will be convoluted to the fluctuations of the barrier due to the tunneling effect.

According to the former TDHFB studies, it can be conjectured the following rule for fusion reactions: In reactions where both fragments are superfluid, the second order fluctuations of the fusion barrier distribution is enhanced compared to similar reactions where at least one of the fragments is not superfluid.
The goal of the present work is to search for the evidence of this effect with a  systematic study of the fusion experimental data. 

Systematic studies of fusion cross section \cite{Siw04,Wan07,Wan17}, usually use a fitting procedure to determine the main parameters of the reaction which are the barrier height, the fusion radius and the width of the barrier. 
This method has a drawback that the final result depends on the choice of the model parametrization.. A new method is proposed and tested in order to determine those three parameters directly from the barrier distribution without assuming a parametrization of the cross sections. 

The paper is organized as follows. The local regression method is tested to reduce the uncertainties on the barrier distribution in Section \ref{Sec:Local_Regression_method}. Then, a benchmark is performed between several methods to determine the fluctuations of the barrier in Section \ref{Sec:Bench}. A systematic analysis of the fluctuations of the barrier is done in Section \ref{Sec:syst}. Finally, the summary is given in Section \ref{sec:summ}

\section{ Local regression method }
\label{Sec:Local_Regression_method}

The fusion barrier distribution is defined as,
	\begin{align}
		D(B) = \left.  \frac1{\pi R_B^2} \frac{ d^2[ E \sigma_{\rm fus}(E) ]  }{dE^2} \right|_{E=B} ,
	\end{align}
with $R_B$ the position of the barrier that is deduced from the normalisation of the barrier distribution. The  second derivative is usually computed with the three-point difference formula, 
	\begin{align}
		 \left. \frac{d^2(E\sigma_{\rm fus}(E))}{dE^2} \right|_{E=E_2} \simeq \frac{E_1 \sigma (E_1) - 2 E_2 \sigma (E_2) + E_3 \sigma (E_3)}{ (\Delta E)^2 }  , \label{eq:3points}
	\end{align}
with $E_1=E_2-\Delta E$ and $E_3=E_2+\Delta E$.
The limitation of this method is the presence of large uncertainties due to the calculation of the second derivative. These uncertainty $\Delta D$ can be estimated at the point $E$ by
$ \Delta D =  \Delta\sigma(E) \sqrt{6} E\sigma(E)/(\Delta E)^2$ \cite{Tim98}.
In practice, to diminish the uncertainties,   the value of $\Delta E$ is increased. 
This produces a smoothing of the barrier distribution. Then, structures in the barrier distribution
smaller than $\Delta E$ will not be visible. It is also necessary in experiments to have a
fixed $\delta E$ step when the center of mass energy varies.

Then $\Delta E$ will be a multiple of the  $\delta E$ value.
In practice, with this method, a part of the information contained in the experimental data is lost because the second derivative at the point $E_2$
is computed from the information of only three points while there can be other experimental points at the 
vicinity of $E_2$ that can bring information on the second derivative.

	From this statement, a new technique to calculate the second derivative using the local regression method is proposed here. The idea is to fit the experimental data around the point at energy $E$ with a polynomial function. The fitting procedure is done with a weight function, 
	\begin{align}
 		 W(E') = \left\{
           	  \begin{array}{cc}
          	    0  & \quad |E'-E|  > L \cr
          	    ( 1 -  (|E'-E|/L)^3 )^3 & \quad |E'-E| \leqslant L
         	    \end{array}             
          	   \right.  ,
 	\end{align}
 with $L$ an adjustable parameter which controls how wide is the window around a point $E$. The parameters $a_i$ of the polynomial function,
	 \begin{align}
		 f_E(x) = \sum_{i=0}^{N} a_i x^i , 
	 \end{align}
are then  adjusted to reproduce the experimental value of $\sigma(E)$. Then, by making this fitting procedure for each window centered on varying energy $E$, the local regression function $F(E)=f_E(E)$ is obtained.  If it is assumed that the cross section varies smoothly in the windows around the energy E, the  function $F(E)$ is expected to be closer to the real $\sigma(E) $ function than the experimental data that contains a statistical uncertainty.

\begin{figure}[htb]
\begin{center}
\includegraphics[width=  \linewidth]{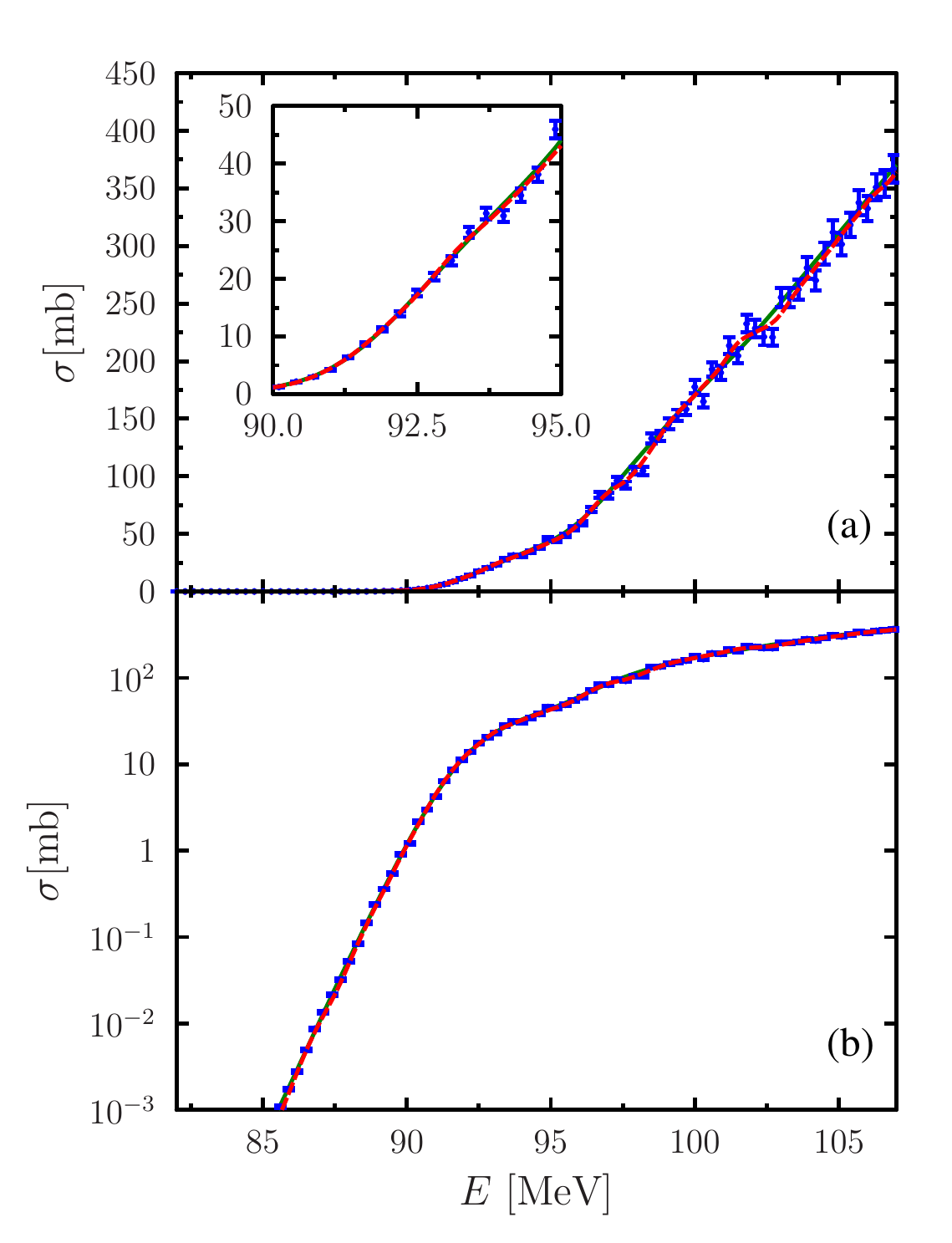}
\end{center}
\caption{ Simulated fusion cross section obtained with the  {\tt CCFULL} program, in linear scale (a) and logarithmic(b). 
The original data are shown with green lines, the blue dots represent the data with a noise and the result of the local regression method $F(E)$ is shown by the red dashed lines.   } 
\label{fig:lin_ccful_cross}
\end{figure}

To test, this method, a fusion cross section is simulated with the  program {\tt CCFULL} \cite{Hag99}. 
The reaction $^{40}$Ca+$^{96}$Zr is computed with a nucleus-nucleus Woods-Saxon potential with a parameter set
$V_0$=87.00 MeV, $r_0$=1.13 fm and $a$=0.7 fm.
 The 3$^-$ collective excitation at energy $E_3$ = 1.89 MeV
of the $^{96}$Zr  are taken into account up to three phonons with a deformation parameter $\beta_3$ =  0.305 
and the 3$^-$ at energy $E_3$=3.7 MeV of the $^{40}$Ca  are taken into account up to three phonons with a deformation 
parameter $\beta_3$=0.43. 
On this data, a random error is added with an amplitude of 5\% and 2\%. Note that in order to describe the reaction $^{40}$Ca+$^{96}$Zr, 
it is necessary to take into account the transfer channel \cite{Ste07,Sca15,Esb16}. Nevertheless, the goal of this calculation is not to realistically describe the fusion barrier distribution of this system but to test the method in the case of a complex barrier which has clear structure effects.

\begin{figure}[htb]
\begin{center}
\includegraphics[width=  \linewidth]{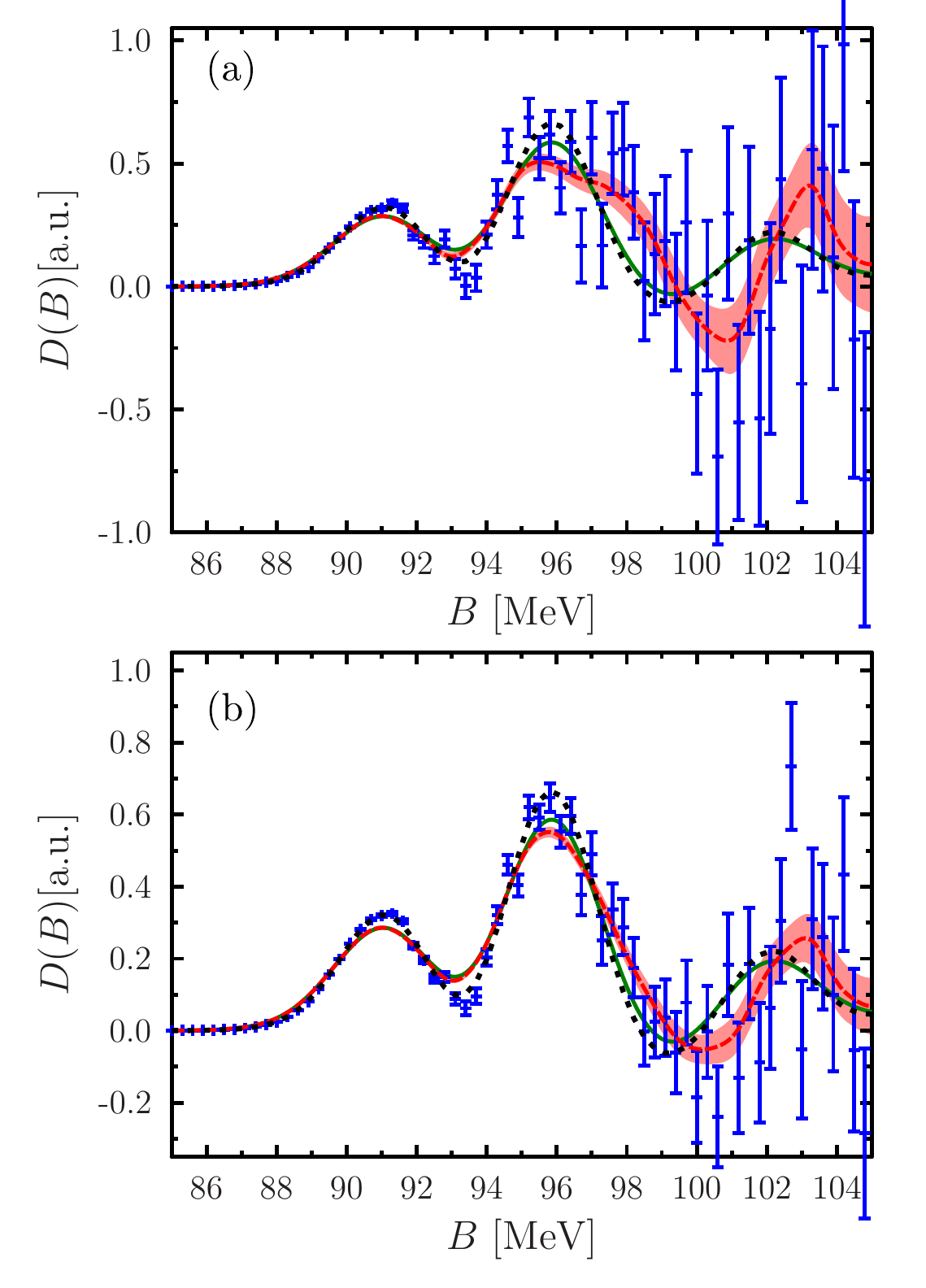}
\end{center}
\caption{  Barrier distribution computed from the original {\tt CCFULL} results with the local regression method (green solid line) and with the three-point formula  (blacked dotted line).
 The results obtained from the data with noise are shown using  the local regression method (red line with colored band) and the  three-point formula (blue points with error bars) . The second derivative is computed with the parameter $\Delta E=$2 MeV.
  An artificial noise of 5\% is applied on (a) and a noise of 2\% on (b).} 
\label{fig:test_barrier}
\end{figure}

The local regression method  with a polynomial function at first order and a parameter $L$=2 MeV 
is then applied to this data and compared to the original cross section of Fig. \ref{fig:lin_ccful_cross}. 
The  function obtained is found to be closer to the original cross section than the simulated experimental points.

From this function, the second derivative is computed with the three-point formula eq. \eqref{eq:3points}.
Note that it is still needed to use a large $\Delta E$ to avoid the overfitting problem.
To estimate the uncertainties, a Monte-Carlo technique is used. 
A set of points $\{\sigma_i\}$ is created, where each point is modified with a random variable $ \sigma_i \rightarrow  \sigma_i + \zeta_i $ with $ \langle \zeta_i \rangle = 0$ and $ \langle \zeta_i^2 \rangle = \delta_i$, with $\delta_i$ the uncertainty on the experimental point (here the artificial error). All the $\zeta_i$ are independent. From this sample, the barrier distribution $D(B)$  is determined. This operation is repeated  $N_{rand}$ times with other random selection.
After $N_{rand}$ samples, the value of $D(B)$ is computed as the average value and the uncertainty as the standard deviation for each point.
In this calculation, the value  $N_{\rm rand}$=100 is chosen. The result with this method is shown on
Fig. \ref{fig:test_barrier} with two artificial noises of 5\% and 2\%. One can see that the local regression 
method is more precise than the direct three-point formula. The error bars are smaller and the average
curve is closer to the exact solution.

Also, in Fig.  \ref{fig:test_barrier} (a), on the region between 100 MeV and 105 MeV, the results of the three-point formula do not bring any information on the barrier. While with the local regression, one can see a barrier at a position close to the real one. The position and the amplitude get closer to the real one when the percentage of error is reduced (See Fig.  \ref{fig:test_barrier} (b)).  Then this method, by reducing the uncertainties allows more fine analysis of the structure of the barrier from experimental cross section data (see for example \cite{Das98,Mon17}).

\begin{figure}[htb]
\begin{center}
\includegraphics[width=  \linewidth]{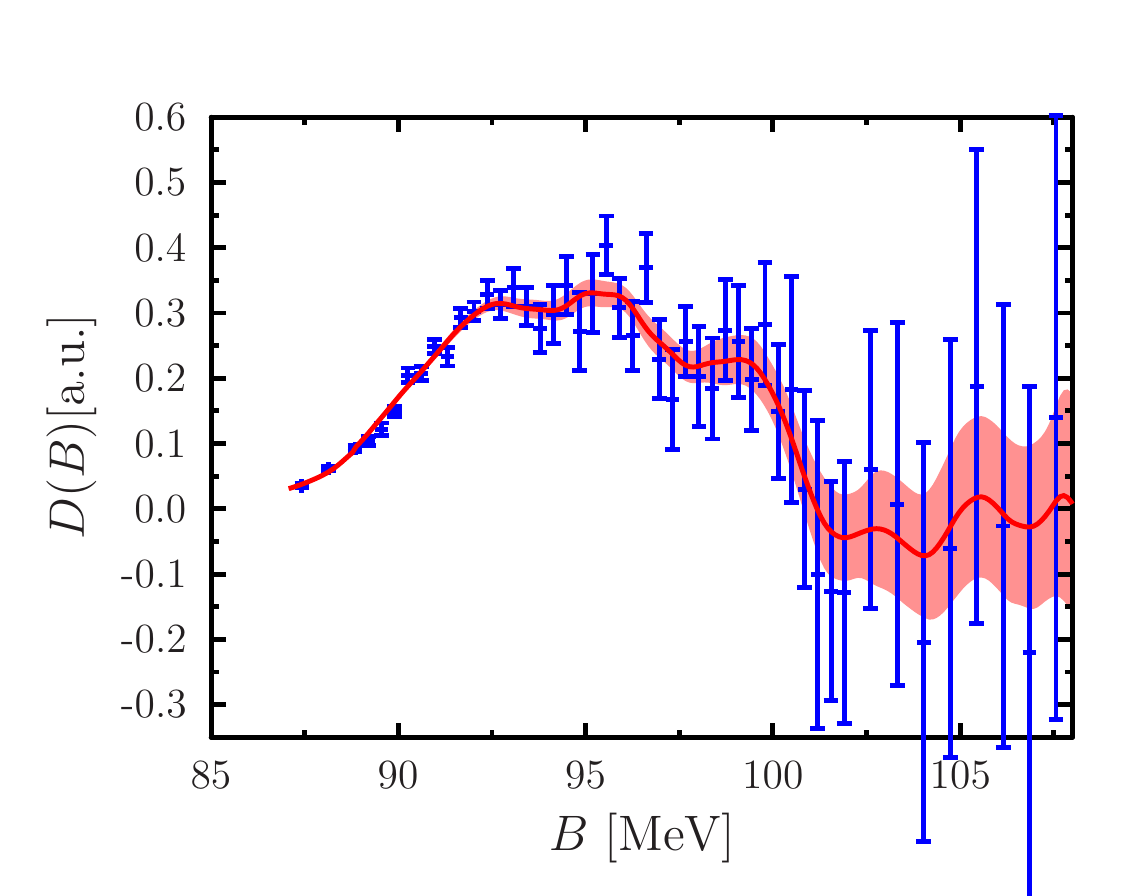}
\end{center}
\caption{  Barrier distribution for the reaction $^{40}$Ca+$^{96}$Zr computed from the experimental cross section \cite{Tim98} with the three-point formula (blue points with error bars) and from the local regression ( red curve and shaded area). The value of $L=\Delta E$= 1.77 MeV is used. } 
\label{fig:bar_comp_exp}
\end{figure}

In Fig. \ref{fig:bar_comp_exp}, this method is tested on the real experimental data \cite{Tim98} of the reaction $^{40}$Ca+$^{96}$Zr. One can see that the three-points formula induces large uncertainties while the local regression method reduces those uncertainties. Another advantage of this method is to provide a continuous function which can be integrated.

\section{Determination of the barrier parameters }

In order to describe the fusion barrier, three parameters are defined, the centroid barrier,
\begin{align}
	B_0 	 &=  \frac{m^{B}_1}{m^{B}_0},  \label{eq:comp_B} 
\end{align}
the fusion radius, defined in order to normalize the barrier distribution,
\begin{align}
	R_B &= \sqrt{ \frac{m^{B}_0}{ \pi } }, \label{eq:comp_R_B}
\end{align}
and the barrier width,
\begin{align}
	\sigma_B 	&= \sqrt{  \frac{m^{B}_2}{m^{B}_0} - \left( \frac{m^{B}_1}{m^{B}_0} \right)^2}. \label{eq:definition_sigmaB}
\end{align}
These three parameters are computed from the moment of the barrier distribution, 
\begin{align}
  m^{B}_n = \int_0^{E_M}   B^n \left. \frac{d^2}{dE^2}\left( \frac{}{} E \sigma(E)  \right) \right|_{E=B}  dB. \label{eq:moment}
\end{align}
$E_M$ is the maximum barrier energy. This formula assumes that above the barrier $E_M$ the barrier distribution is zero.

\label{Sec:Bench}
\subsection{Calculation from the barrier distribution}

\begin{figure}[htb]
\begin{center}
\includegraphics[width=  \linewidth]{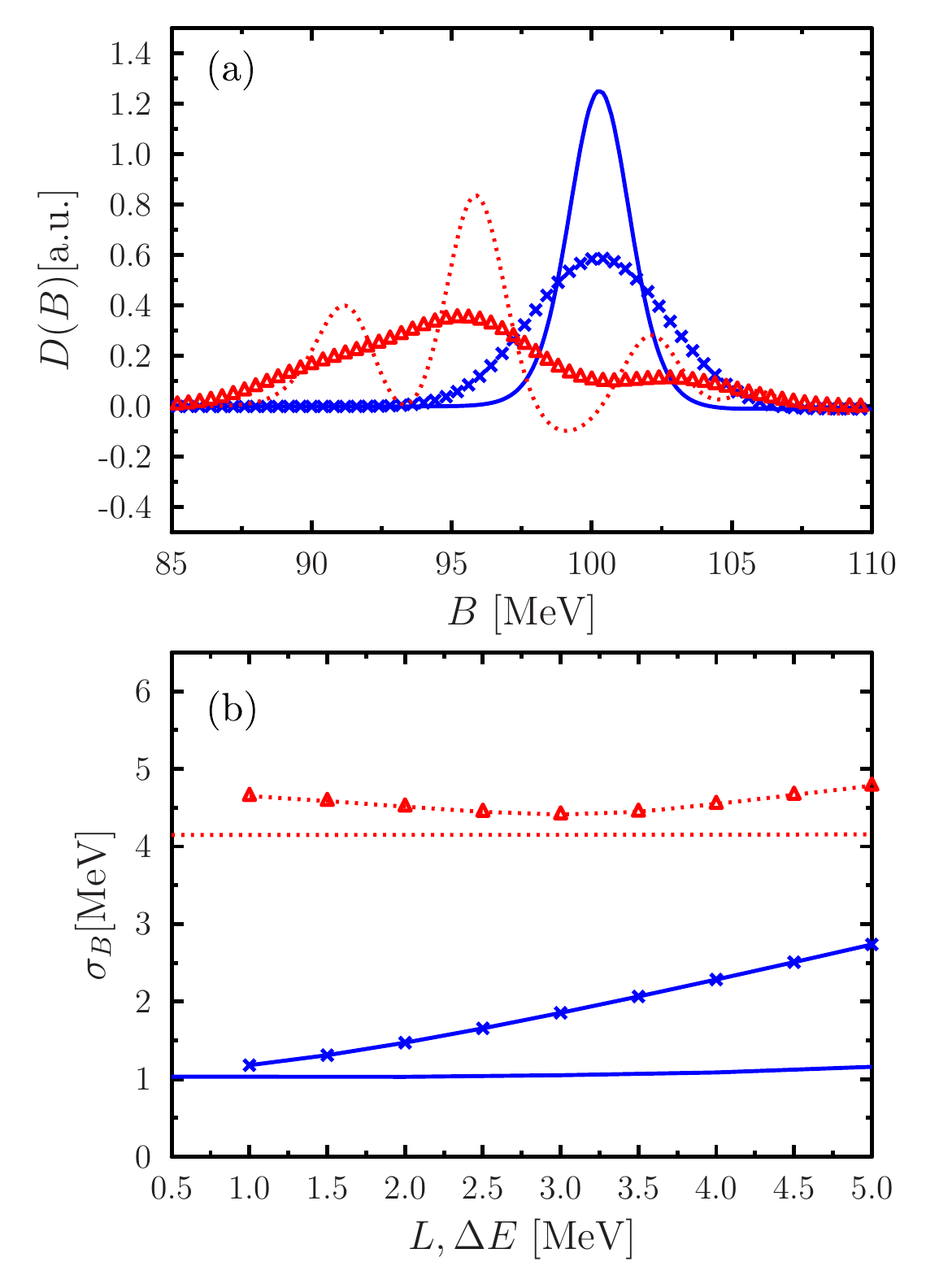}
\end{center}
\caption{  (a) Barrier distribution computed by the local regression method for the simulated data
 with parameter $\Delta E$ = $L$ = 1 MeV (solid and dotted lines) and 4 MeV (triangle and crosses markers) 
 computed from the {\tt CCFULL} calculation with (red dotted line and triangles) and without the collective 
 excitations (blue solid line and crosses). 
(b) Fluctuations of the barrier distribution from the {\tt CCFULL} calculation
 with (red triangles) and without the collective excitations (blue crosses) as a function of the three-point  derivative parameter $\Delta E$. A comparison is made with the integration method (eq. \eqref{eq:sigm_int}) with (red dotted line) and without collective excitations (blue solid line) as a function of $L$. } 
\label{fig:sigma_fct_delta_E}
\end{figure}

In order to determine the fluctuations of the barrier, the standard deviation of the barrier (eq. \eqref{eq:definition_sigmaB})
is computed, with the integration made only with the points that have a positive value of $D(B)$.

The difficulty of this method is that the result depends on the parameter $\Delta E$ used to compute the barrier. 
To show this phenomenon, the effect of the parameter $\Delta E$  on the barrier distribution is shown in Fig. \ref{fig:sigma_fct_delta_E}a.
Two test cases are shown, the first one is the same cross section as Sec \ref{Sec:Local_Regression_method} computed with the 
collective 3$^-$ excitations that create structures on  the barrier distribution and a calculation without any collective excitation. 
The second barrier is almost Gaussian and has  small fluctuations. When  the value of the $\Delta E$ parameter increases,
the barrier distribution is spread, then  the value of $\sigma_B$ increase. 

The obtained value of $\sigma_B$ as a function of $\Delta E$ is shown in Fig. \ref{fig:sigma_fct_delta_E}b. The value needed is the asymptotic value when  $\Delta E$ tends to zero, which is difficult to attend in practice.
It is then not possible to determine the correct value of  $\sigma_B$ without being dependent on the parameter $\Delta E$.
Note that in practice, it is also difficult to determine the maximum energy $E_M$.

\subsection{Integral method}

\begin{figure}[htb]
\begin{center}
\includegraphics[width=  \linewidth]{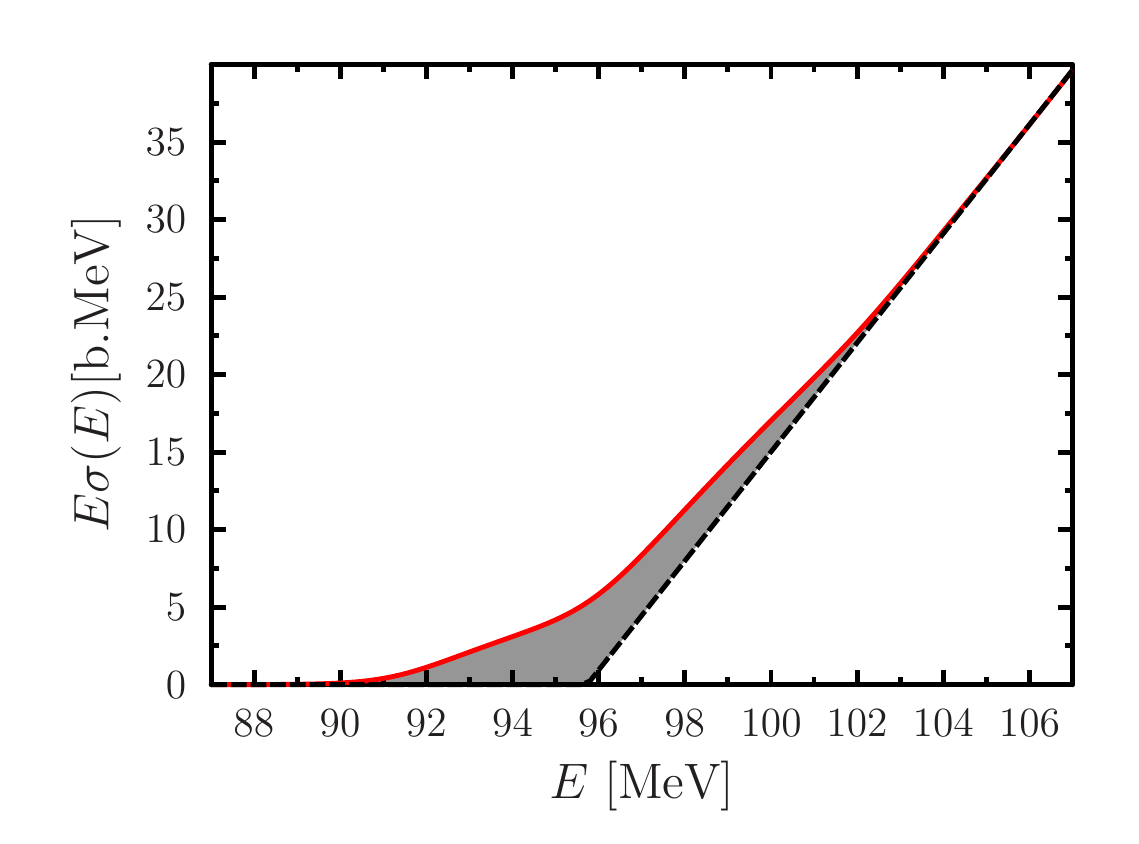}
\end{center}
\caption{$^{40}$Ca+$^{96}$Zr {\tt CCFULL}  fusion cross section multiplied by the energy (red solid line). The function $g(E)$ is shown with a dashed black line. The shaded area represents the integral of eq. \eqref{eq:sigm_int}.} 
\label{fig:meth_integr}
\end{figure}

One can avoid the calculation of the second derivative and then avoid the problem of convolution found in the previous section by using partial integration on  eq. \eqref{eq:moment},
\begin{align}
m^{B}_0 &=  \left. \frac{d}{dE}\left( \frac{}{} E \sigma(E)  \right) \right|_{E=E_M}, \\
m^{B}_1 &= E_M( m^{B}_0 - \sigma(E_M)), \\
m^{B}_2 &= E_M^2 ( m^{B}_0 - 2 \sigma(E_M)) + 2 \int_0^{E_M} E \sigma(E) dE.
\end{align}
From which simple expressions of the main parameters of the barrier are deduced, 
\begin{align}
R_B^2  &= \frac1{\pi} \left. \frac{d}{dE}\left( \frac{}{} E \sigma(E)  \right) \right|_{E=E_M}, \\
B_0 		   &= E_M \left( 1 - \frac{ \sigma(E_M) }{ \pi R_B^2} \right), \\
\sigma_B^2 &= \frac{2}{ \pi R_B^2 } \int_0^{E_M} \left(  E \sigma(E) - g(E) \right) dE, \label{eq:sigm_int}
\end{align}
 with
\begin{align} 
 		 g(E) = \left\{
           	  \begin{array}{cc}
          	    0  & \quad E \leqslant B_0 \cr
          	     \pi R_B^2 ( E -B_0  )  & \quad E>B_0 \cr
         	    \end{array}             
          	   \right.  .
\end{align}

This method requires computing the derivative of the fusion cross section at the energy $E_M$ and one integral. The integral is computed from the local regression function $F(E)$. In practice,  the function $g(E)$ is first adjusted to the experimental curve (see Fig. \ref{fig:meth_integr}) around the point $E_M$, and then the integral of Eq. \eqref{eq:sigm_int} is computed from the local regression function $F(E)$. Note that this method is close to the one of Ref. \cite{Das04} to compute the centroid of the barrier distribution $B_0$.

Using this method on the {\tt CCFULL}  cross section, the values of the barrier fluctuations are $\sigma_B$ = 4.18 MeV and $\sigma_B$ = 1.03 MeV respectively with and without excitations. Those values are very stable with the $L$ parameter as shown in Fig.  \ref{fig:sigma_fct_delta_E}b. As one can expect from the Fig. \ref{fig:meth_integr}, the area between the two curves is very dependent on the slope of the g(E) function. 
This method is then limited to the experimental data where the slope above the barrier can be well determined. Note that the fitting method should also be very dependent on the slope above the barrier, but, it will not be explicit in the fitting procedure. In case of data without a clear slope above the barrier, the fitting procedure will extrapolate from the cross section data below the barrier. This extrapolation, if the barrier is more complicated than the fitting function will not be accurate.

\begin{figure*}[htb]
\begin{center}
\includegraphics[width=  \linewidth]{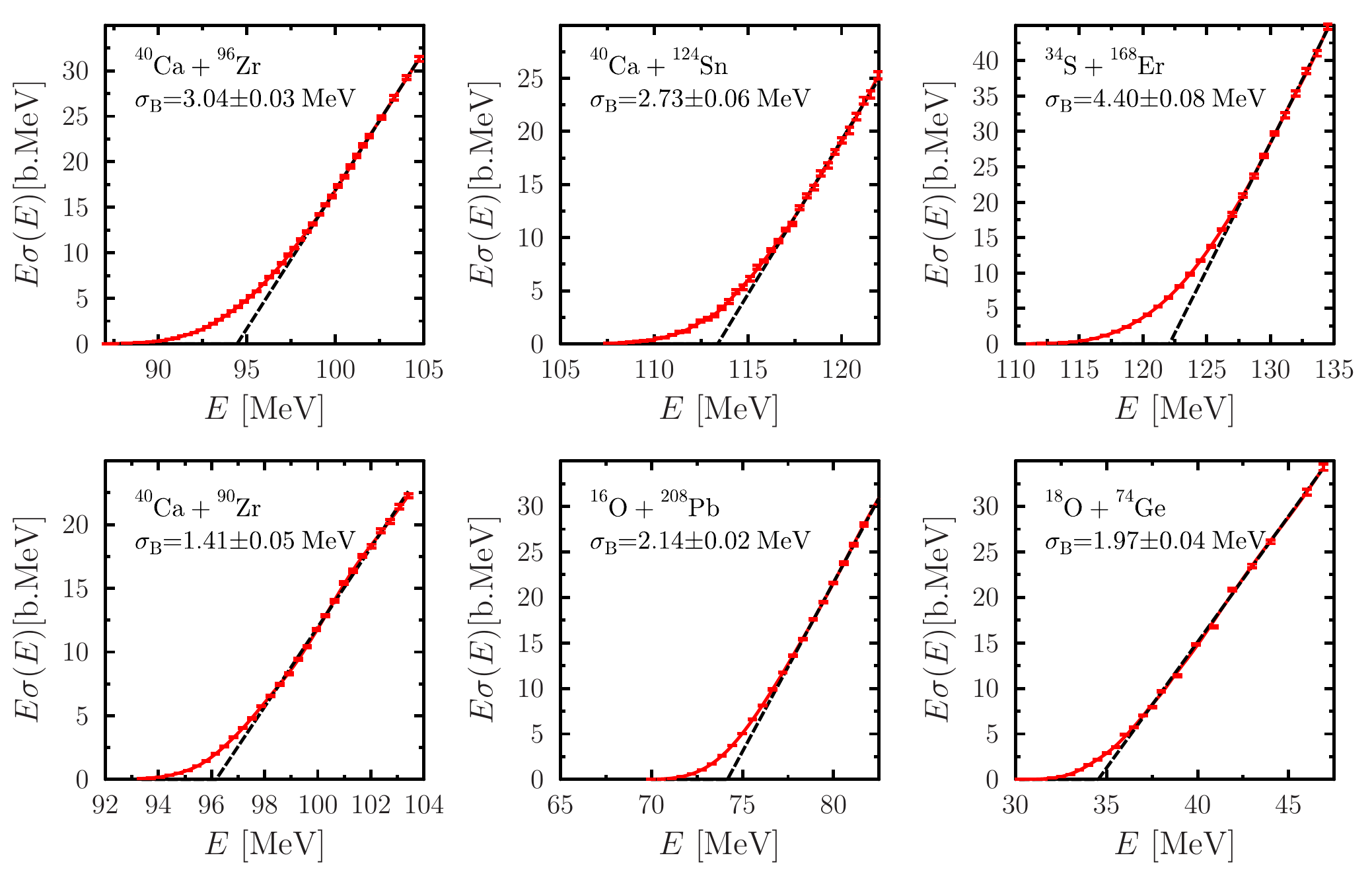}
\end{center}
\caption{ Example of the application of the integration method for several reactions. The black dashed line represents the $g(E)$ function and the red cross the experimental data. The experimental data are taken from Refs. \cite{Tim98,Sca00,Mor00,Mor99,Jia12}. } 
\label{fig:cross}
\end{figure*}

Several examples of applications of this method are shown in Fig. \ref{fig:cross}. To determine the uncertainties the same Monte-Carlo method than for the barrier is used. For each of those examples, the linear $g(E)$ function can be adjusted to the experimental data without ambiguity. In this panel of 6 cross sections, the uncertainties on the values of the fluctuations of the barrier vary from 1\% in the case where the quality of the experimental cross section is very good ($^{40}$Ca+$^{96}$Zr) to 4\% where the number of points is less important and the uncertainties larger ($^{40}$Ca+$^{90}$Zr).

\subsection{Fitting procedure}

\begin{figure}[htb]
\begin{center}
\includegraphics[width=  \linewidth]{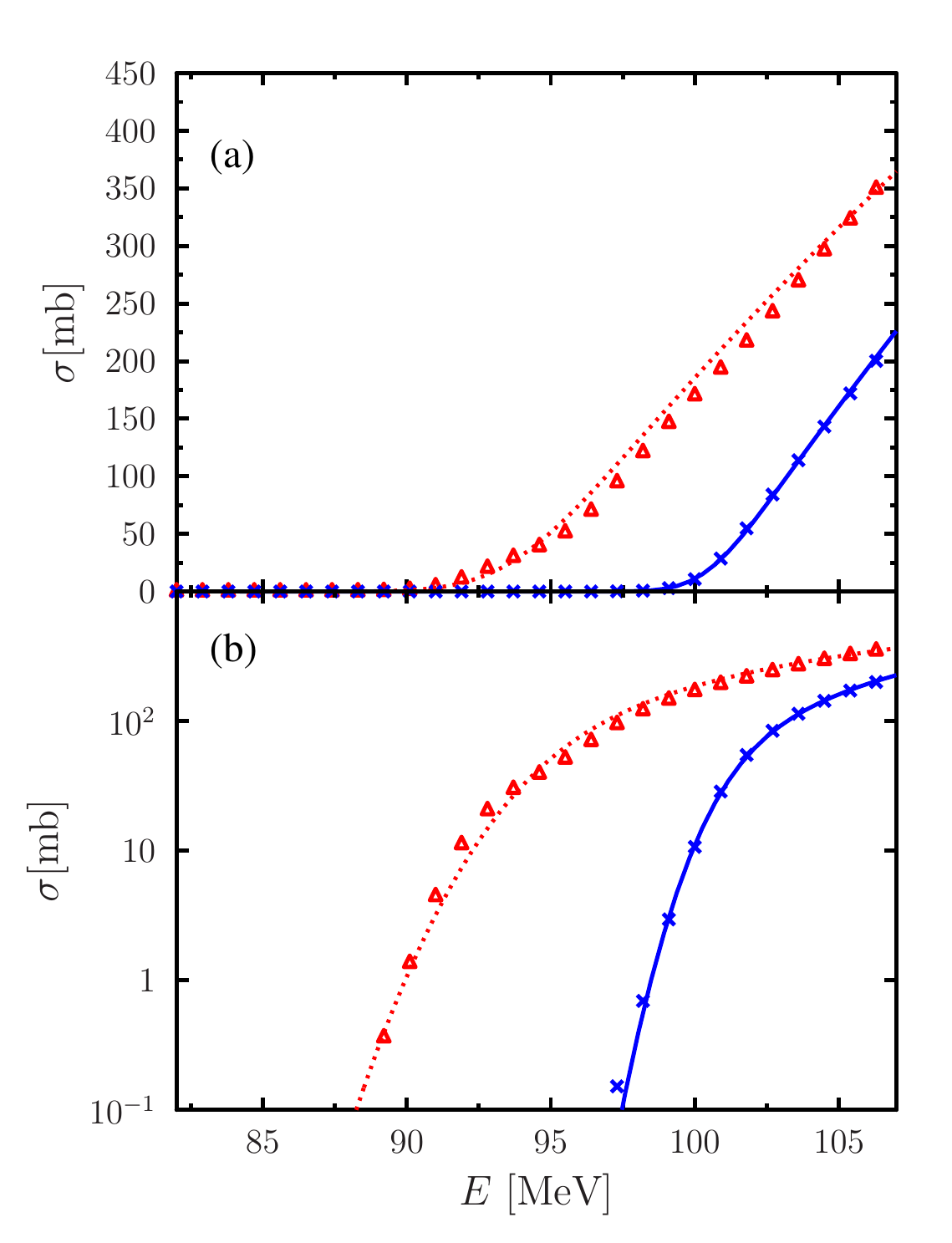}
\end{center}
\caption{$^{40}$Ca+$^{96}$Zr {\tt CCFULL}  fusion cross section calculation with (red triangles) and without (blue crosses) collective excitation. The function eq. \eqref{eq:fct_fit_swi} is  adjusted to those cross sections and shown respectively, with a red dotted line and a blue solid line.  } 
\label{fig:comp_fit_Swi}
\end{figure}

Another method to determine the parameters of the barrier is to fit the experimental data with a parametrization of fusion cross section \cite{Swi05},
\begin{align}
	\sigma_{\rm fus} = \pi R_B^2  \frac{\sigma_B}{E\sqrt{2\pi}} [X\sqrt{\pi} (1+ {\rm erf} X )  + \exp(-X^2)],\label{eq:fct_fit_swi}
\end{align}
with $X=\frac{E-B_{0}}{\sqrt{2}\sigma_B}$.
The parametrization of the fusion cross section corresponds to a Gaussian barrier distribution
with standard deviation $\sigma_B$. The parameters of this function are adjusted on the fusion
cross section obtained with the {\tt CCFULL}  program. In the case with the excitations, the parameters
are $R_B$=11.47 fm,  $B_0$=93.66 MeV and $\sigma_B$=2.08 MeV. In the case
where the excitations are not taken into account $R_B$=12.85 fm,  $B_0$=100.4 MeV
and $\sigma_B$=1.18 MeV.

In the second case, the value of $\sigma_B$ is very close to the one in Fig. \ref{fig:sigma_fct_delta_E} in the limit of $\Delta E$ small. In the case of a single barrier almost gaussian, the two methods give the same result. But with the cross section generated with structure effects, the barrier is no more Gaussian. Then the fit underestimates a lot the barrier fluctuations, $\sigma_B$=2.08 MeV instead of about 4.5 MeV with the direct calculation. 

To go beyond this approach, the fusion cross section is fitted with a sum of two functions of Eq. \eqref{eq:fct_fit_swi} which is equivalent to assume that the barrier is composed of a sum of two Gaussians. Then the barrier width is determined by Eq. \eqref{eq:definition_sigmaB}.
With this method, the barrier fluctuations are of 3.44 MeV.
 This result is closer to the correct value, but still underestimate the real fluctuations of the barrier. Note that the interesting method of the Bayesian spectral deconvolution \cite{Hag16} could improve the present fitting procedure, but seems to be too complex to be used for a systematic analysis.

\section{Systematic analysis}
\label{Sec:syst}

\begin{figure}[htb]
\begin{center}
\includegraphics[width=  \linewidth]{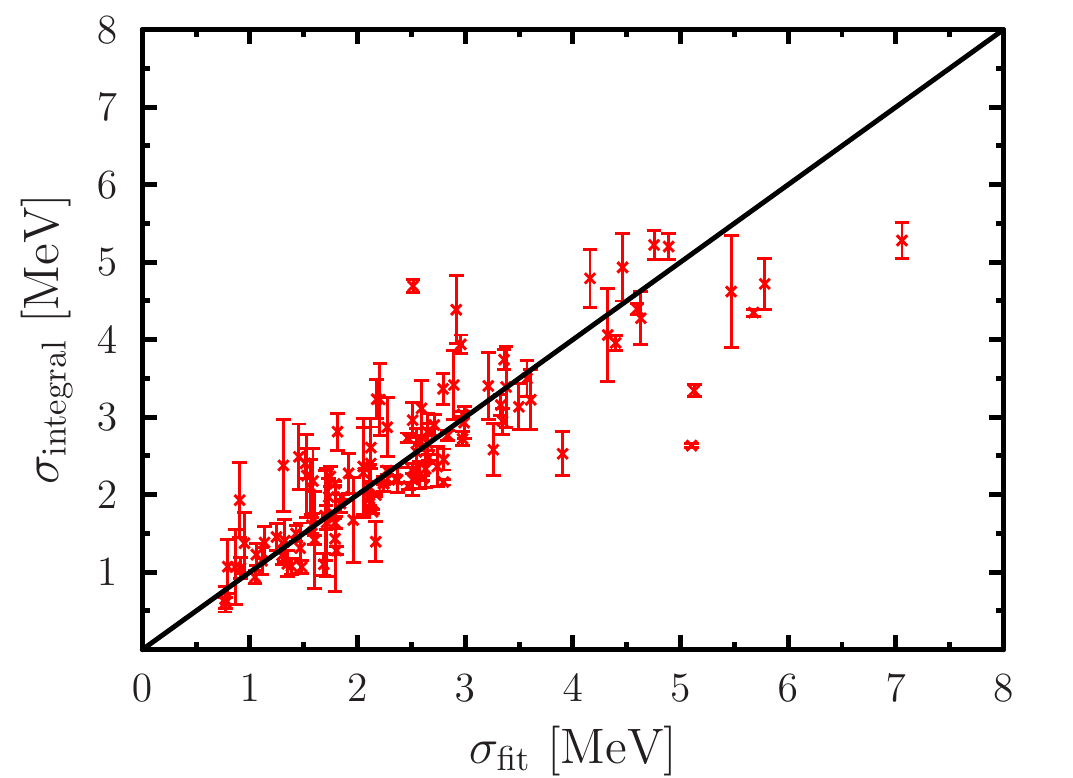}
\end{center}
\caption{ Fluctuations of the barrier $\sigma_B$ determined with the integral method as a function of the fit method.   } 
\label{fig:comp_fit_integral}
\end{figure}

The two methods (fitting procedure with two Gaussians and integral method) have been systematically applied to a large number of experimental data from the database \cite{nrv}. 115 reactions have been selected on those data for which the slope above the barrier can be reasonably well determined. The main selection has been done on the uncertainties of the results. Only systems for which the uncertainties on the value of $\sigma_B$ is lower than 0.75 MeV have been analyzed.
A comparison between the results obtained by both methods is shown in Fig. \ref{fig:comp_fit_integral}. A good agreement is found between the two methods. For 73\% of the reactions, the two methods are giving results with a difference of less than 0.5 MeV.

In order to analyze the data, I define the parameter $S$ that reflects the superfluidity of the reaction.
For one reaction, this parameter is computed as follow: starting with $S=0$; if the $N_1$ AND $N_2$ are non magic $S$ is changed to 1; then, if the $Z_1$ AND $Z_2$ are non magic $S$ is incremented by 1. $N_1$ and $N_2$ are the neutron numbers of the two nuclei. $Z_1$ and $Z_2$ are the proton numbers. The magic number  taken here are $\{ 8, 20, 28, 50, 82, 126 \}$.

The value of $S$ can take three values 0, 1 and 2. If it is assumed that only the non-magic number nuclei are superfluid, then, for systems with $S=0$, no increase of the fluctuations of the barrier is expected. While with $S=1$ or $S=2$ it can be expected that the superfluidity will increase the fluctuations of the barrier and that the effect will be larger with $S=2$ where neutrons and protons of each fragment are supposed to be in the superfluid phase.

\begin{figure}[htb]
\begin{center}
\includegraphics[width=  \linewidth]{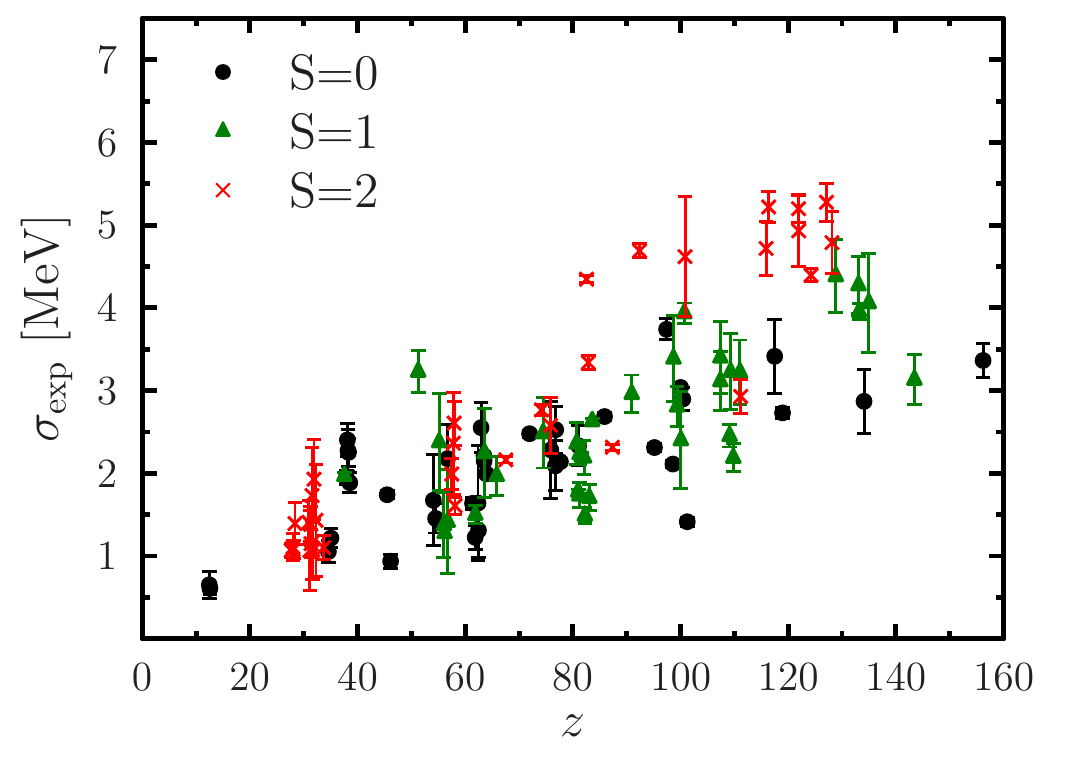}
\end{center}
\caption{ Fluctuations of the barrier $\sigma_B$ determined with the integral method as a function of the $z$ parameter.   } 
\label{fig:sigma_fct_z}
\end{figure}

In order to not mix the superfluid effects with the fusion hindrance, only systems with $Z_1Z_2<$1500 are selected. A naive comparison of the different systems with different values of $S$ is shown in Fig. \ref{fig:sigma_fct_z}. Where the obtained $\sigma_B$ with the integral method is shown as a function of the parameter $z=\frac{Z_1 Z_2}{A_1^{1/3}+A_2^{1/3}}$.  For reactions with $z$ below 80, no effects are seen and all the reactions have small fluctuations of about 2 MeV.  For systems with $z>80$ three groups can be identified, those with small $\sigma_B$ around 2 or 3 MeV, those systems are mainly $S=0$, those with $\sigma_B$ around 3 or 4 MeV which are mainly $S$=1 and the last group around 5 MeV is mainly composed of systems with $S=2$. 

Then, this first result corresponds to the expected result with the tendency $\sigma_B^{S=2}>\sigma_B^{S=1}>\sigma_B^{S=0}$. Nevertheless, this analysis neglects all the other effects that play a role in the determination of the fluctuation. In particular, the deformation that is also related to the magicity of the initial fragments.

\begin{figure}[htb]
\begin{center}
\includegraphics[width=  \linewidth]{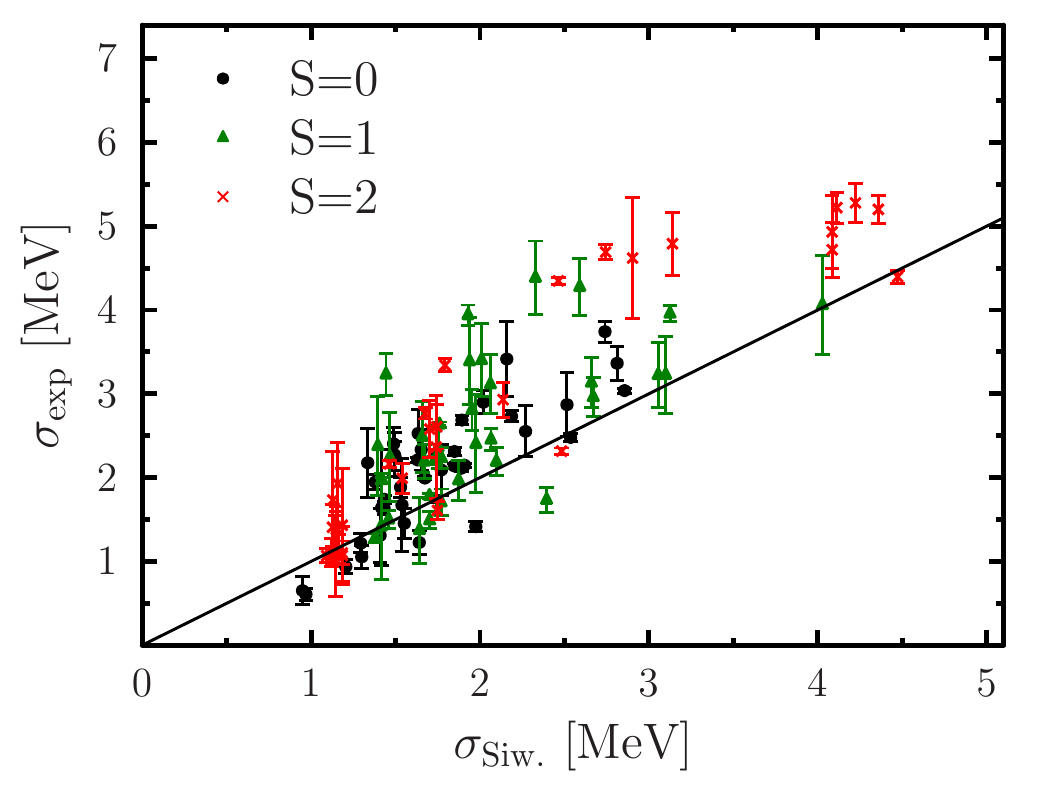}
\end{center}
\caption{ Fluctuations of the barrier $\sigma_B$ determined with the integral method as a function of the estimated barrier from Ref. \cite{Siw04}.   } 
\label{fig:comp_sigma}
\end{figure}

In order to take into account those effects,  the estimate $\sigma_B$  is computed from the model of ref. \cite{Siw04}. This model takes into account three sources of fluctuations of the barrier, (i) the tunneling effect, (ii) the static deformation, (iii) the vibration. Then, the total width of the barrier is computed as the convolution of these effects for each fragment (1) and (2),
\begin{align}
(\sigma_B^{\rm Siw})^2 &=  {\sigma_{\rm Tunnel}}^2 + \sigma_{\rm Static}(1)^2 +  \sigma_{\rm Static}(2)^2 \nonumber \\
&+ \sigma_{\rm Vib.}(1)^2 +  \sigma_{\rm Vib.}(2)^2 . \label{eq:sigma_siw}
\end{align}  
The formula of each of the terms are given in Ref.  \cite{Siw04}. This model is empirical and has several parameters adjusted on the experimental data on a large number of systems. For each reaction, the total width of the barrier is computed only from the input $A_1$, $A_2$, $Z_1$, $Z_2$ and the $\beta_2$ of each of the fragments. The $\beta_2$ values are taken from the M\"oller table \cite{Mol95}.

Because this last model does not take into account the effect of the superfluidity, it is expected for the systems with S=1 or S=2 that 
the empirical model will under-estimate the fluctuations of the barrier ($\sigma_{\rm exp.} > \sigma_{\rm Siw.}$).
A comparison between the experimental values of the fluctuations of the barrier and the obtained values from the empirical model is made in Fig. \ref{fig:comp_sigma}.

 From this comparison, one can observe the following. i) On average, the Siwek-Wilczynska model underestimates the width of the barrier. This is due to the tendency of the fitting procedure used in Ref. \cite{Siw04} to underestimate the barrier width and to the larger number of reactions studied here.
 ii) The experimental fluctuations of the barrier are in the range of 0 to 6 MeV. There is no system that is compatible with very large fluctuations of the order of 10 MeV. iii) A clear effect of the superfluidity is found in several reactions with $S=$1 or 2 which are found to have a larger barrier width than the expected value from the Siwek-Wilczynska model and from the general trend of systems with $S$=0.

\begin{table}[h]
\caption{ Systems with $S$=1 or 2 where an enhancement of the fluctuations of the barrier more than 1 MeV is found. The values of the $\sigma$ are given in MeV. The type of the experiment evaporated residue (EvR) or fusion-fission (FF) done is shown in the last column. }
\centering \begin{tabular}{|c|c|c|c|c|c|}
\hline\hline
Reaction & $\;S\:$ &$\sigma_{\rm Siw.}$  &   $\sigma^{\rm integ.}_{\rm exp.}$ &   $\;$ Ref. $\;$ & exp. \\ 
 \hline
$^{40}$Ar+$^{144}$Sm	&  1  &  2.31  &  4.39 $\pm$ 0.44  &  \cite{Rei85}   &  EvR+FF \\
$^{32}$S+$^{138}$Ba  	&  1  &  2.06  &  3.11 $\pm$ 0.35  &  \cite{Gil95}    &  EvR+FF \\
$^{40}$Ar+$^{122}$Sn  	&  1  &  1.94  &  3.41 $\pm$ 0.44  &  \cite{Rei85}   &  EvR+FF \\
$^{32}$S+$^{120}$Sn  	&  1  &  1.94  &  3.39 $\pm$ 0.52  &  \cite{Tri01}    &  EvR \\
$^{58}$Ni+$^{94}$Zr  	&  1  &  2.59  &  4.28 $\pm$ 0.34  &  \cite{Sca91}  &  EvR \\
$^{58}$Ni+$^{60}$Ni  	&  1  &  1.87  &  3.94 $\pm$ 0.12  &  \cite{Ste95b} &  EvR \\
$^{19}$F+$^{93}$Nb  	&  1  &  1.44  &  3.23 $\pm$ 0.25  &  \cite{Pra96}  &  EvR \\
$^{40}$Ar+$^{154}$Sm  	&  2  &  4.27  &  5.28 $\pm$ 0.23  &  \cite{Rei85}  &  EvR+FF \\
$^{40}$Ar+$^{148}$Sm  	&  2  &  3.15  &  4.79 $\pm$ 0.37  &  \cite{Rei85}  &  EvR+FF \\
$^{32}$S+$^{110}$Pd  	&  2  &  2.65  &  4.69 $\pm$ 0.09  &  \cite{Ste95}  &  EvR \\
$^{40}$Ar+$^{110}$Pd  	&  2  &  2.90  &  4.62 $\pm$ 0.73  &  \cite{Jah82}  &  EvR \\
$^{32}$S+$^{96}$Zr  	&  2  &  2.46  &  4.35 $\pm$ 0.05  &  \cite{Zha10}  &  EvR \\
$^{32}$S+$^{94}$Zr  	&  2  &  1.79  &  3.34 $\pm$ 0.08  &  \cite{Jia14}   &  EvR \\
$^{28}$Si+$^{178}$Hf 	&  2  &  4.11  &  5.22 $\pm$ 0.18  &  \cite{But02}   &  EvR+FF \\
$^{28}$Si+$^{92}$Zr  	&  2  &  1.68  &  2.77 $\pm$ 0.07  &  \cite{New01} &  EvR \\
 \hline\hline
\end{tabular}
\label{Tab:S2_value_inc}
\end{table}

The Tab. \ref{Tab:S2_value_inc}  presents systems with S=1 or 2  that have larger fluctuations of the barrier than the estimated value from the model.   The table is given here, in order to guide the future microscopic applications of TDHFB or other models that aim to quantitatively reproduce  the effect of the superfluidity on the barrier.

\begin{figure}[htb]
\begin{center}
\includegraphics[width=  \linewidth]{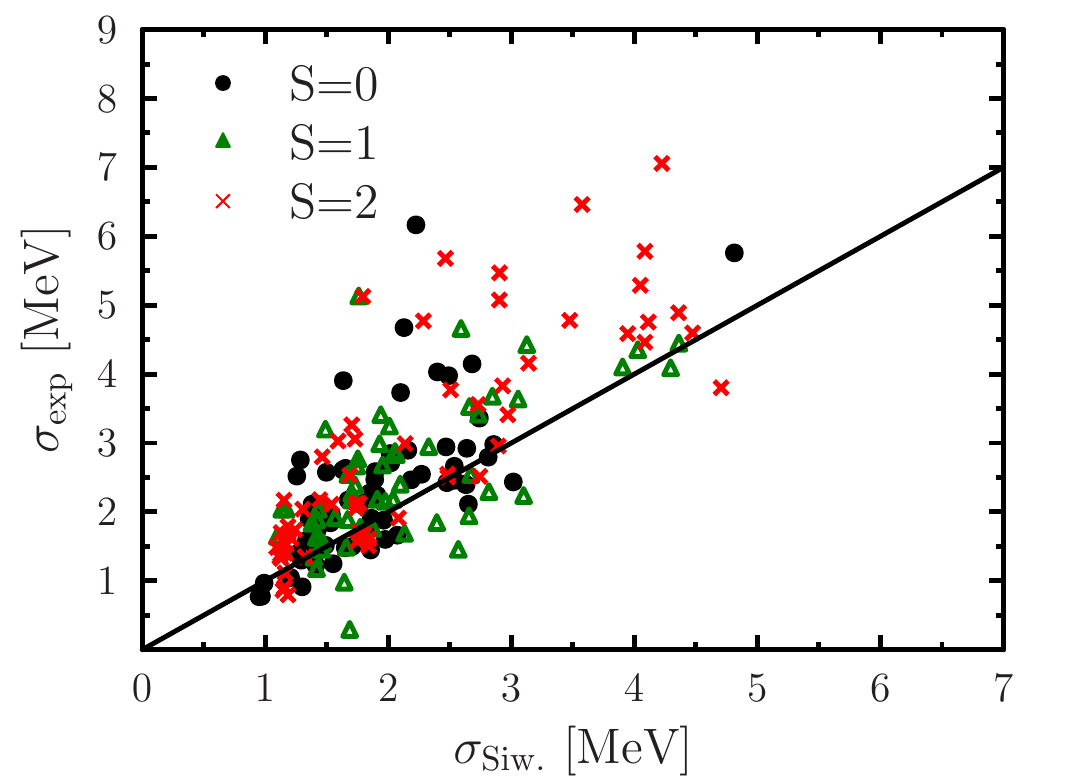}
\end{center}
\caption{ Same as Fig. \ref{fig:comp_sigma} with the $\sigma_{\exp}$ determined with the fitting procedure.  } 
\label{fig:comp_sigma_fit_alpha}
\end{figure}

In order to confirm the results of the Fig. \ref{fig:comp_sigma}, the same analysis is done with the fitting method in Fig. \ref{fig:comp_sigma_fit_alpha}. The results of the fitting method are expected to be of lower quality, but the method is more tolerant of the quality and quantity of points in the experimental data. Then, this systematic analysis includes 194 reactions. Those results are shown to confirm the enhancement of the fluctuations of the barrier for systems where $S$=1 or 2. Note that, the points with $S$=0 which presents a large width of the barrier are not present in the Fig. \ref{fig:comp_sigma} because they have too large uncertainties.

\begin{figure}[htb]
\begin{center}
\includegraphics[width=  \linewidth]{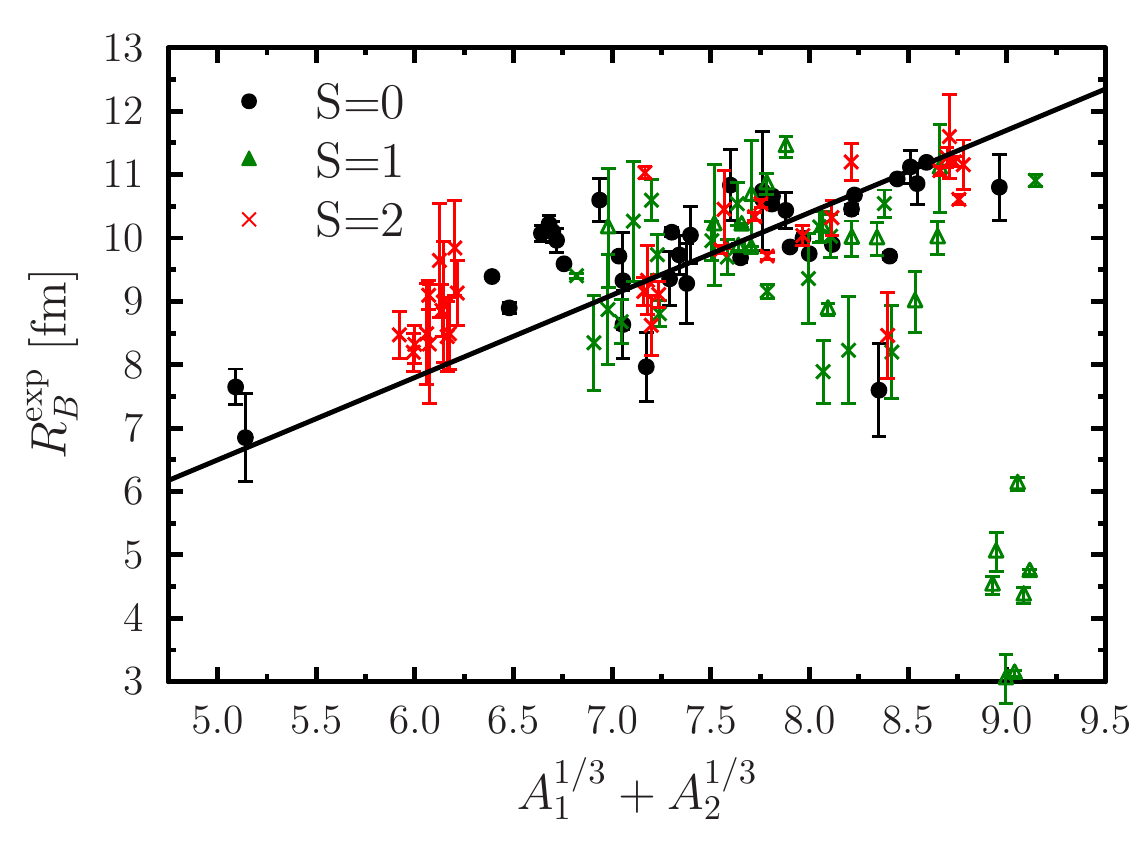}
\end{center}
\caption{  Experimental fusion radius computed as eq. \eqref{eq:comp_R_B}.  The solid line represent the function $R_B=1.30 (A_1^{1/3}+A_2^{1/3})$. } 
\label{fig:R_B_fctA13}
\end{figure}

\begin{figure}[htb]
\begin{center}
\includegraphics[width= \linewidth]{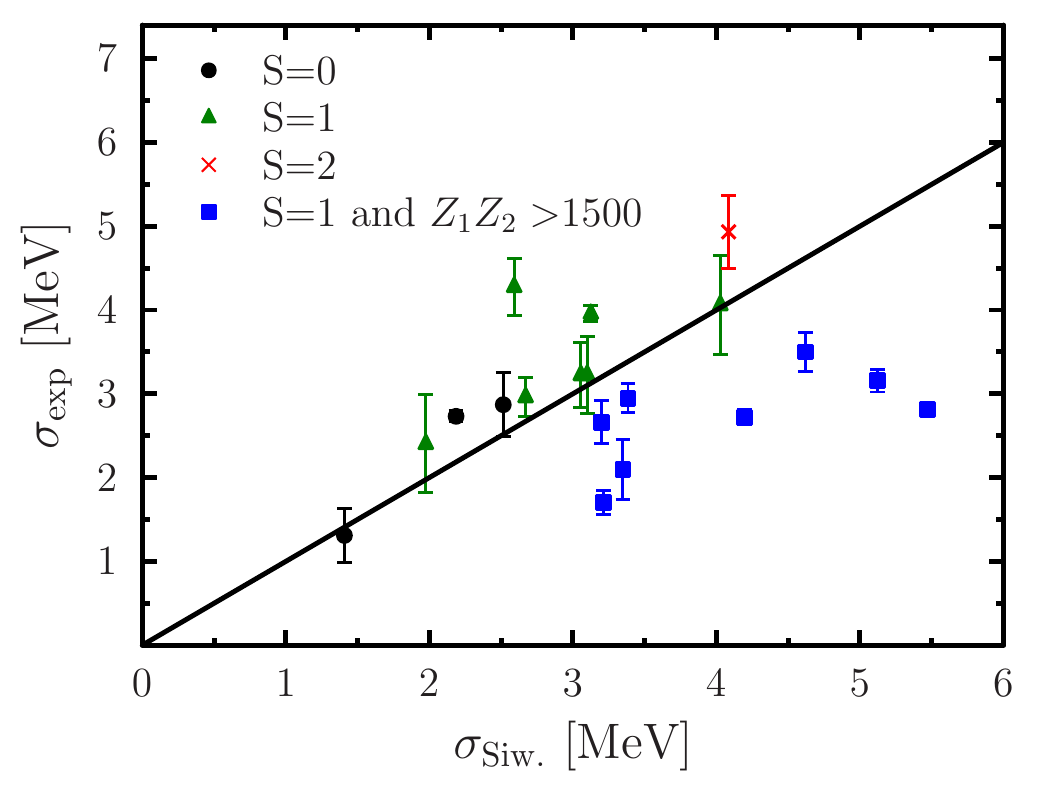}
\end{center}
\caption{ Same as Fig. \ref{fig:comp_sigma} with a selection of systems for which $R_B^{\rm exp.} < 1.3 (A_1^{1/3}+A_2^{1/3}) - 1$ fm. } 
\label{fig:comp_sigma_select2}
\end{figure}

To finish this empirical analysis,  the effect of the superfluidity on the fusion radius and on the centroid of the barrier distribution is investigated. 
In fig. \ref{fig:comp_sigma_fit_alpha}, the fusion radii of all the selected reactions, including the systems where it is expected an effect of the fusion hindrance ($Z_1Z_2>$1500) are shown.  Those last reactions do not follow the general trend $R_B\simeq 1.3 (A_1^{1/3}+A_2^{1/3}) $ and present a small radius in the range $3<R_B<6.3$ Fm due to the fusion hindrance. 
Those data are composed of systems close to Z1$\simeq$ Z2 $\simeq$ 40   and the experimental data from Refs. \cite{Bec84,Kel86,Rei85b} are done detecting only the evaporated residue. The width of the barrier is shown in this selection of reactions on the Fig. \ref{fig:comp_sigma_select2} by blue squares. For those reactions, the eq. \eqref{eq:sigma_siw} clearly overestimate the fluctuations of the barrier. 
It is difficult to say if this reduction of the fluctuations of the width is due to hindrance effect or due to the absence of the fusion-fission detection in the experiments. In future analysis, it would be interesting to do a similar systematic analysis for heavy systems where hindrance plays a role. Note that shell structure effects have already been shown on fusion hindrance \cite{Sat02} and competition with quasi-fission \cite{Sim12}.

This last hypothesis is corroborated by another selection of reactions which are found to have a reduced fusion radius ($R_B^{\rm exp.} < 1.3 (A_1^{1/3}+A_2^{1/3}) - 1$ fm). They have different values of $Z_1Z_2$ in a range from 700 to 1200. Then, it is not expected any fusion hindrance effects. Those systems are analyzed here from experimental data \cite{Ste95,Sca91,Bec83,Bec82,Sca00,Rei85b,Tri01} which are also done detecting only the evaporated residue. The Fig. \ref{fig:comp_sigma_select2} shows that those systems mainly do not present the expected enhancement of the barrier width due to the superfluidity. 

To understand this effect due to the lack of detection of the fusion-fission fragments, the method is tested on the experimental data for the $^{58}$Ni+$^{132}$Sn reaction, where evaporated residue and fusion-fission data are available \cite{Koh11}. If only the data of the evaporated residue is considered, the width found is $\sigma_{\rm exp}$=2.77 MeV and the fusion radius $R_B^{\rm exp}$=7.46 fm. While with the complete data (evaporated residue and fusion-fission) it is found $\sigma_{\rm exp}$=3.36 MeV and $R_B^{\rm exp}$=10.8 fm. This result shows that one can expect the real fluctuations barrier to be higher for the reactions shown in Fig. \ref{fig:comp_sigma_select2}. Then, it can explain why for those reactions with $S$=1 there is no visible effect of the superfluidity.

\begin{figure}[htb]
\begin{center}
\includegraphics[width= \linewidth]{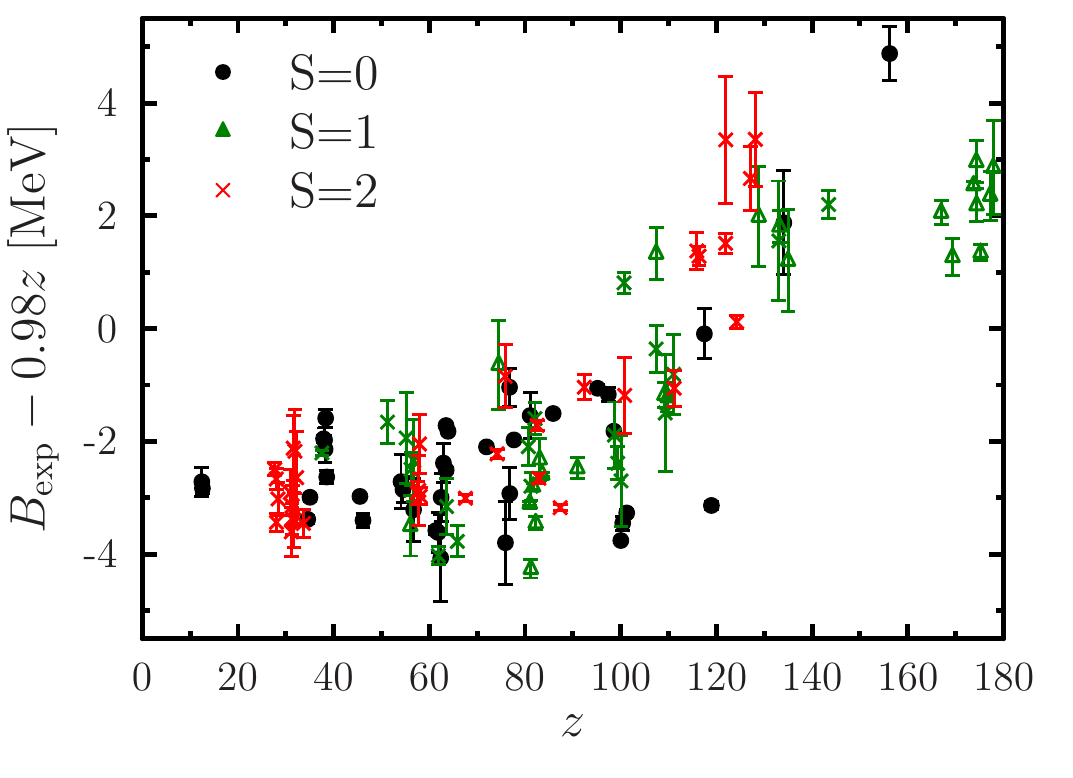}
\end{center}
\caption{ Deviation of the experimental centroid of the fusion barrier distribution computed as \eqref{eq:comp_B} from the general trend $0.98 z$ as a function of $z$.  } 
\label{fig:B_fct_z}
\end{figure}

Concerning the effect of the superfluidity on the fusion radius, apart from the reactions discussed before, there is no apparent correlation between the fusion radius and the superfluid number $S$. This is also the case for the barrier height. As shown in Fig. \ref{fig:B_fct_z}, the deviation of the barrier from a simple linear function $B=0.98 z$ does not present a correlation with the superfluidity.

\section{Summary} 

\label{sec:summ}
In this work, I develop new methods to determine the width of the fusion barrier distribution. I first use the local regression method to compute the fusion barrier. This method is more precise than the three-point formula. It presents smaller uncertainties and allows more fine analysis of the barrier structure. Nevertheless, this method is not able to accurately determine  the fluctuations of the barrier. 

I propose a second method that requires only the integration of the fusion cross section. This method is more robust than the fitting procedure because it does not assume any shape of the barrier distribution.

This method is applied to 115 fusion reactions and compared to a model that does not include the expected effect of the superfluidity. An enhancement of the fluctuations of the barrier of about 1 MeV is found in several reactions between superfluid nuclei. This result proves that the effect predicted by TDHFB calculation is real. Nevertheless, this empirical result is in contradiction with the idea of a very strong effect of the superfluidity in the fusion barrier. No effect of the superfluidity is found on the barrier height and on the fusion radius.

In addition to this results, a list of reactions between non-magic nuclei which present an enhancement of the fluctuations of the barrier distribution is provided. Futur microscopic calculations of the fusion barrier should be applied to this list in order to reach a better comprehension of the effect of the superfluidity on the fusion barrier. It would be also interesting to investigate the effect of the pairing on the effective coordinate mass \cite{Hag07,Uma14}, that could also be a source of correlations between the superfluidity and the width of the barrier.

\begin{acknowledgments}

\end{acknowledgments}

I would like to thanks K. Hagino, Y. Tanimura, H. Sagawa, G. Wlazlowski and P. Magierski for interesting discussions and T. Nakatsukasa for his careful reading of the manuscript.

\end{document}